\documentclass[useAMS,usenatbib]{mn2e}
\usepackage{graphicx}
\usepackage{natbib}
\usepackage{subfigure}

\usepackage{subfigure}
\newcommand{\el}[1]{\raisebox{0ex}{\scriptsize #1}}

\bibpunct{(}{)}{;}{a}{}{,}
\title[Monte Carlo simulations of H$_{2}$ formation on stochastically heated grains]{Monte Carlo simulations of H$_{2}$ formation on stochastically heated grains}

\author[H.~M.~Cuppen, O.~Morata and Eric Herbst]{H. M. Cuppen$^{1}$, O.~Morata$^{1}$ and Eric Herbst$^{2}$\\
$^{1}$ Department of Physics, The Ohio State University, Columbus, OH 43210, USA\\
$^{2}$ Departments of Physics, Astronomy, and Chemistry, The Ohio State University, Columbus, OH 43210, USA\\
}

\begin{document}
\maketitle

\begin{abstract}
Continuous-time, random-walk Monte Carlo simulations of H$_{2}$ formation on grains have been performed for surfaces that are stochastically heated by photons. We have assumed diffuse cloud conditions and used a variety of grains of varying roughness and size based on olivine. The simulations were performed at different optical depths.  We  confirmed that small grains (\(r\leq 0.02\) \(\mu\)m) have low modal temperatures with strong fluctuations, which have a large effect on the efficiency of the formation of molecular hydrogen. The grain size distribution highly favours small grains and therefore H$_2$ formation on these particles makes a large contribution to the overall formation rate for all but the roughest surfaces. We find that at $A_{V}=0$ only the roughest surfaces can produce the required amount of molecular hydrogen, but by $A_{V}=1$, smoother surfaces are possible alternatives. 
{Use of a larger value for the evaporation energy of atomic hydrogen, but one still consistent with experiment, allows smoother surfaces to produce more H$_{2}$.}

\end{abstract}


\section{Introduction}
Although molecular hydrogen is abundant in diffuse and dense regions of the neutral interstellar medium (ISM), the detailed mechanism of its formation is still not completely clear. At the low gas-phase temperatures typical of the interstellar medium, a gas-phase formation is not possible, and H$_{2}$ is therefore formed on the surfaces of grains. An interpretation of experimental studies \citep{Pirronello:1997,Pirronello:1997a,Pirronello:1999} indicates that the surface temperature range over which efficient H$_{2}$ formation occurs is very small for olivine (6-10 K) and for amorphous carbon (13-17 K) \citep{Katz:1999} under diffuse-cloud conditions. The grain temperature for midsize grains in unshielded regions is however around 20 K. This means that molecular hydrogen cannot be formed efficiently on such grains assuming that the olivine surfaces used in the laboratory are a realistic representation of interstellar grains. The analysis of \cite{Katz:1999} considered, however, only a single barrier for diffusion between binding sites and a single low binding energy between physisorbed H atoms and the surface. In previous papers \citep{Chang:2005, Cuppen:H2}, we showed that the introduction of sites of different binding energy leads to an increase of the temperature range for efficient H$_2$ formation. The different sites are the result of a difference in local environment either due to the amorphous character of the grain or to the  topological structure of the grain; i.e., surface roughness. These results are in agreement with the findings of \cite{Cazaux:2004}, who introduced extra chemisorption sites and made a distinction between H and D atoms. They found a larger temperature range depending on the width and the height of the barrier between sites of physisorption and chemisorption.

In this paper, we will study the formation of molecular hydrogen using similar  continuous-time, random-walk Monte Carlo simulations as in our previous papers \citep{Chang:2005, Cuppen:H2} while grains are heated by the interstellar radiation field. The photons hitting a grain give it a short heat impulse, resulting in temperature fluctuations especially for small grains. These temperature fluctuations have received considerable attention in the literature \citep{Greenberg:1974,Purcell:1976,Aannestad:1979,Tabak:1987,Draine:2001}, where they are discussed with a varying degree of detail. While most of the studies simply use the heat capacity to calculate the temperature changes, \cite{Draine:2001} looked at the heating and cooling of small silicon clusters and PAHs by considering the infrared emission of the grains, taking into account the infrared bands of the grain material. Since our main goal is to study molecular hydrogen formation, we will use a relatively simple model for the heating and cooling. As we will show, this model is capable of reproducing the detailed method to a large extent.

Since both the heating and the recombination of H atoms into H$_{2}$ are stochastic processes, the Monte Carlo technique is an ideal method to study how both processes influence each other. During the simulations, the individual H atoms are followed as they land on the grain surface, undergo their random walk, and evaporate. If two atoms arrive at the same position on the grain, they react to form molecular hydrogen. As these processes occur, photons are hitting the grain, leading to a pulsed temperature increase and subsequent decrease for the smaller grains. The radiation field and the absorption and emission coefficients used in the simulations are explained in Section \ref{rad field}. The times at which the possible events - hopping, evaporation, and deposition of H atoms, and heating of the grain - occur are randomly determined using the rates of the processes and a random number. The details of the Monte Carlo procedure are discussed in \cite{Chang:2005}. We  briefly summarise them in Section \ref{Model}, which also explains the heating and cooling of the grain in more detail. Section \ref{tempfluc} contains our results for the fluctuations in temperature, while Section \ref{temp vs eff} shows their effect on the formation efficiency of H$_2$ as well as the effects of grain temperature vs. grain size and immediate grain desorption following formation of H$_2$. Our results are put into the context of the ISM in Section \ref{ISM}. Finally, our conclusions are discussed in Section \ref{conclusions}.

\begin{figure*}
\begin{center}
\includegraphics[width=0.95\textwidth]{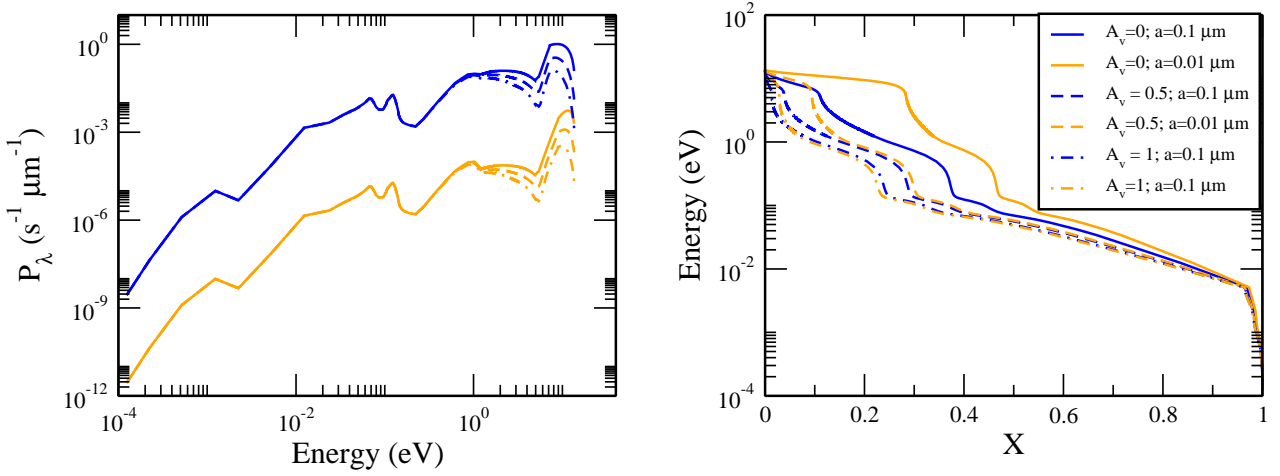}
\end{center}
\caption{(left) The absorbed photon flux in photons per second per wavelength interval and (right) the relation between the random number and the photon energy for different grain sizes, \(r=0.01\) and 0.1 $\mu$m, and at different visual extinctions, \(A_V=0\), 0.5 and 1 (see legend, which applies to both panels).}
\label{dist}
\end{figure*}

\section{The photon flux}
\label{rad field} 
The rate of photons absorbed by a grain depends on its position in the cloud and on its optical properties. We assume the distribution of photons absorbed by the grain per second per wavelength interval as a function of wavelength to be given by
\begin{equation}
P_\lambda = \pi r^2 I_\lambda Q_{abs}(\lambda) D_\lambda,
\label{P_l}
\end{equation}
with \(r\) the radius of the grain, \(I_\lambda\) the interstellar radiation field in photons per unit area per second per unit wavelength, \(Q_{abs}\) the wavelength-dependent absorption coefficient,  and \(D_\lambda\) the reduction factor due to attenuation of the radiation field by dust. 

For the interstellar radiation field \(I_\lambda\) we use different expressions in different regimes. In the high photon energy regime, with wavelengths 91.2-250 nm, the approximation of \cite{Sternberg:1988} is used. The lower photon energy regime, with wavelengths 250 nm-1 cm,  is divided into four different regions using the expressions from \cite{Zucconi:2001}. 

At low photon energies, the absorption coefficient, \(Q_{abs}\), is approximated using 
\begin{equation}
Q_{abs}(\lambda) \approx \frac{8\pi r}{\lambda} {\rm Im}\left(\frac{m^2-1}{m^2+2}\right)
\label{Qabs}
\end{equation}
\citep{Bohren:1983}, where \(m\) is the complex refractive index:
\begin{equation}
m \equiv n + ki.
\end{equation}
For the optical constants \(n\) and \(k\), we used tabulated values for olivine from \cite{Ossenkopf:1992}. Since Eq.~(\ref{Qabs}) is only valid for \(\frac{2\pi r}{\lambda} \ll 1\) we only used this approximation for \(Q_{abs}\) for \(\lambda \geq 1000\) nm. For smaller wavelengths, the $r$ dependence of \(Q_{abs}\) becomes different. Here we used a set of empirical expressions to approximate the absorption coefficients found by \cite{Draine:1984} and \cite{Draine:1985}. They showed the absorption coefficients for silicates and carbonaceous material to be significantly different over a large energy range. We only used the silicate results here. 

For the shielding of the radiation field, we followed \cite{Draine:1996}, who derived the attenuation factor, $D_\lambda$:
\begin{equation}
D_\lambda=\exp\left(- 3.7 \frac{A_\lambda}{A_{\el{1000\AA}}}A_V\right),
\end{equation}
based on the attenuation at 1000 \AA. Since we preferred to use \(A_V\) instead of $A_{\el{1000\AA}}$ we took \(\frac{A_{\el{1000\AA}}}{A_V} = 4.70\) from \cite{Whittet:2003} to arrive at the alternative expression
\begin{equation}
D_\lambda=\exp\left(- 0.8 \frac{A_\lambda}{A_V}A_V\right).
\end{equation}
We then used the table of \(\frac{A_\lambda}{A_V}\) values in \cite{Mathis:1990} and \cite{Whittet:2003}.

The left panel of Figure \ref{dist} gives the total  flux of absorbed photons  \(P_\lambda\) in photons per second for grains of two different sizes and at different visual extinctions as a function of energy. The shape of the curves for different grain sizes is the same for low energies because of the constant \(r^3\) dependence stemming from the linear dependence of \(Q_{abs}(\lambda)\). This relationship  no longer holds in the high energy regime where the curve shapes are different for different grain sizes. This regime can be divided in two ranges: \(P_\lambda\) depends on more than \(r^3\) from approximately 1-7 eV while the dependency is less strong for energies higher than 7 eV. 
The graph further shows that dust attenuation mainly affects the high energy regime. 

\section{The surfaces and the model}
\label{Model}
In our previous paper \citep{Cuppen:H2}, we performed our Monte Carlo simulations of the recombination of hydrogen on different surfaces of varying surface roughness. These surfaces were generated using a separate Monte Carlo simulation program starting from a flat surface with a square grid to represent binding sites on a grain. The resulting surfaces were characterised by a matrix indicating the topological roughness of each lattice grid. We used four different surfaces: surface (a) is completely flat while surfaces (b), (c), and (d) have protrusions and ''holes''.  In addition, surface (c) has several features known as  islands and surface (d) is very rough with height differences of several monolayers and no distinct shape for the protrusions. If a hydrogen atom lands on the irregular surface, its local hopping and evaporation energy are calculated based on its number of horizontal `grain' neighbours, \(i\) (see Fig.~2 in \cite{Cuppen:H2}). For every horizontal neighbour, there is an extra interaction energy \(E_L\). This method effectively results in surfaces with a maximum of five different binding sites. The present paper uses the same concept to construct the three surfaces used in the new simulations. Surface I has only one type of site and is therefore flat. Surface II has a roughness comparable to surface (c) in \cite{Cuppen:H2}, but we only distinguish two types of sites.  The first type of site has zero ($i=0$) horizontal neighbours while the second type, which can have  one or more horizontal neighbours according to the method used to generate the surface, is still assumed to have only one neighbour ($i=1$). Surface III derives from surface (d) in \cite{Cuppen:H2} and has five different types of sites, with 0 to 4 horizontal neighbours. 

Figure \ref{surf} shows representations of  the non-flat surfaces II and III. Note that the colours used in this figure indicate the different sites, in contrast with Figure 1 in \cite{Cuppen:H2} where the colour coding is used to indicate the different heights. For surface II, the black sites correspond to \(i=0\) and the white sites to \(i=1\), whilst for Surface III, the lighter the colour, the greater the number of horizontal neighbours \(i\). The hopping ($E_b$) and evaporation ($E_D$) energies of these sites, labelled by the number of neighbours \(i\), are given by
\begin{equation}
E_{b,i} = E_{b} + iE_{L},
\end{equation}
\begin{equation}
E_{D,i} = E_{D} + iE_{L},
\end{equation}
respectively. The energies \(E_{b}\), \(E_{D}\), and \(E_L\) used here for H and H$_2$ are given in Table \ref{E} for the different surfaces along with the total number of different sites, \(i_{tot}\) and the distribution of these different sites \(n_i\). The lateral bond, $E_L$, is chosen to be 40\% of the evaporation barrier, \(E_{D}\). Our previous paper \citep{Cuppen:H2} showed that the recombination efficiency for surface III is close to unity for temperatures around 20 K. With this choice of the lateral bond, we can distinguish three cases: zero, low, and high efficiency for the surface temperatures of interest in the absence of fluctuations.

\begin{figure}
\begin{center}
\includegraphics[width=0.3\textwidth]{s2s.eps}
\includegraphics[width=0.3\textwidth]{s5s.eps}
\end{center}
\caption{The distribution of the olivine surface sites for (top) \(i_{tot}=2\) and (bottom) \(i_{tot}=5\). The black area indicates the sites with \(i=0\) and the lighter colours are for increasing \(i\).}
\label{surf}
\end{figure}

\begin{table}
\caption{The parameters characterising the different surfaces used. }
\label{E}
\begin{tabular}{lrrr}
\hline
                       & \multicolumn{1}{c}{I} & \multicolumn{1}{c}{II} & \multicolumn{1}{c}{III}\\
\hline
\(E_{b,\el{H}}\) (K)\(^1\)   & 287                   & 287                   & 287                   \\
\(E_{b,\el{H}_2}\) (K)\(^2\) & 242                   & 242                   & 242                   \\
\(E_{D,\el{H}}\)  (K)\(^1\)  & 373                   & 373                   & 373                   \\
\(E_{D,\el{H}_2}\) (K)\(^1\) & 315                   & 315                   & 315                   \\
\(E_{L,\el{H}}\)  (K)\(^3\)  &                       & 149                   & 149                   \\
\(E_{L,\el{H}_2}\) (K)\(^3\) &                       & 126                   & 126                   \\
\(\nu\) (s\(^{-1}\))\(^4\)   & 10\(^{12}\)           & 10\(^{12}\)           & 10\(^{12}\)           \\
\(i_{tot}\)                  & 1                     & 2                     & 5                     \\
\(n_0\)                      & 1                     & 0.910                 & 0.540                 \\
\(n_1\)                      &                       & 0.090                 & 0.258                 \\
\(n_2\)                      &                       &                       & 0.126                 \\
\(n_3\)                      &                       &                       & 0.052                 \\
\(n_4\)                      &                       &                       & 0.024                 \\
\hline
\end{tabular}
\\
\(^1\) taken from \cite{Katz:1999}.\\
\(^2\) 0.77\(E_{D,\el{H}_2}\), where 0.77 is the ratio between \(E_{b,\el{H}}\) and \(E_{D,\el{H}}\) \citep{Ruffle:2000}.\\
\(^3\) \(E_{L} = 0.4E_{D} \)\\ 
\(^4\) \cite{Biham:2001}
\end{table}

Each Monte Carlo cycle starts by checking which event will occur next. This is done by comparing the times at which all future events occur: heating of the grain, cooling of the grain, deposition of an H atom, or hopping or evaporation of a particle on the surface. All these events and their corresponding times are discussed in the following three sections. We start by explaining the events involved in the recombination of molecular hydrogen while Sections \ref{heat} and \ref{cool} contain discussions of the heating and the cooling of the grain, respectively.

\subsection{Desorption, hopping, and evaporation of H and H$_2$}
The deposition rate of H atoms (s$^{-1}$ site$^{-1}$) is calculated from the gas abundance using
\begin{equation}
R_{dep} = \frac{v n(\rm{H})}{4\rho}
\end{equation}
where \(n(\rm{H})\) is the concentration of atomic hydrogen, \(\rho\) is the surface site density, and \(v\) is the average velocity of the H atoms in the gas:
\begin{equation}
v = \sqrt{\frac{8 kT_{gas}}{\pi m}},
\end{equation}
with \(T_{gas}\) the temperature of the gas and \(m\) the atomic mass. We use \(n(\rm{H})=10^{2}\) cm\(^{-3}\), $T_{gas}$ = 60 K,  and \(\rho = 2 \times 10^{14}\) cm\(^{-2}\) \citep{Biham:2001}. The time between two deposition events is determined using a random number \(X\) between 0 and 1, and the deposition rate:
\begin{equation}
\Delta t_{dep} = -\frac{\ln(X)}{R_{dep}N_{s}},
\label{t_dep}
\end{equation}
where \(N_{s}\) is the number of sites on the grain surface.

When a deposition event occurs, the deposition site is determined using another random number. If the atom lands on a bare site of the grain, it simply lands there and the time at which the particle will do its next event is determined. This is done in a way similar to determining \(\Delta t_{dep}\):
\begin{equation}
t_{event} = -\frac{\ln(X)}{R_{hop}+R_{eva}} + t,
\label{t_event}
\end{equation}
where \(t\) is the current time in the simulation and \(R_{hop}\) and \(R_{eva}\) are respectively, the hopping rate and the evaporation rate of the particle. These two rates are given by 
\begin{equation}
R_{hop,i} = \nu\exp\left(-\frac{E_{b,i}}{T}\right)
\end{equation}
and
\begin{equation}
R_{eva,i} = \nu\exp\left(-\frac{E_{D,i}}{T}\right)
\end{equation}
respectively, where \(T\) is the temperature of the dust and \(\nu\) is a trial frequency here chosen to be 10\(^{12}\) s\(^{-1}\) (Table \ref{E}). A second random number is used to determine which event, hopping or evaporation, will occur at that time.

If the particle (H) lands on top of another particle (H or H$_2$), it is checked if a reaction between the landing species and species already there can occur. If so, the old particle is replaced by the new molecule, otherwise the incoming particle remains on top of the old one with hopping and evaporation rates, $R_{hop,0}$ and $R_{eva,0}$, respectively. In both cases, the time of the next event for the new or newly formed particle is determined. We assume initially that all H\(_2\) molecules that are formed stay on the surface upon reaction and later evaporate. This corresponds to the parameter choice \(\mu = 1\) introduced by \cite{Katz:1999}{, where $\mu$ indicates the fraction of molecules that remains on the surface immediately following reaction.}

\subsection{Heating of the grain}
\label{heat}
We  consider photons in the wavelength range 91.2 nm-10 mm. The rate (s\(^{-1}\)) of photons absorbed by a grain is 
\begin{equation}
\label{R_photon}
R_{photon} = \int_{91.2\rm{nm}}^{10\rm{mm}}P_\lambda \textrm{d}\lambda
\end{equation}
from which the time between two photon hits can be determined in a manner similar to Eqs.~(\ref{t_dep}) and (\ref{t_event}).
If the grain absorbs a photon, the energy $E$ of the photon is picked from the distribution in Eq.~(\ref{P_l}) using a random number. The right panel in Figure \ref{dist} indicates the relation between the random number, \(X_\lambda\), and the photon energy. This figure is obtained using the relation
\begin{equation}
X_\lambda = \frac{\int_{91.2\rm{nm}}^{\lambda}P_\lambda \textrm{d}\lambda}{R_{photon}}
\end{equation}
which gives a value between 0 and 1 for a given wavelength \(\lambda\). This value is different for different grain sizes and different visual extinctions. The \(r\) dependence is due to the \(r\) dependence in the absorption coefficient.  

The new temperature is then obtained by solving
\begin{equation}
E = \int_{T_{old}}^{T_{new}} c(T)\textrm{d}T
\end{equation}
for \(T_{new}\), where \(c(T)\) is the heat capacity. According to \cite{Aannestad:1979}, \cite{Draine:2001}, and \cite{Purcell:1976}, olivine can be approximated by a Debye solid with a Debye temperature of 500 K. The low temperature heat capacity in the Debye model is
\begin{equation}
c(T\rightarrow 0) = \frac{12\pi^4}{5} N k \left( \frac{T}{T_D}\right)^3,
\label{Debye}
\end{equation}
with \(N\) the number of atoms in the grain. If we assume a density of $3.32$ g cm$^{-3}$ and a molecular mass of 153.3 g mol\(^{-1}\), the heat capacity in eV K\(^{-1}\) becomes  
\begin{equation}
c(T) = 61.38 r^3T^3,
\label{c(T)}
\end{equation}
with \(r\) in $\mu$m. 

Every time the grain temperature changes, the time of the next event for each particle on the surface needs to be adjusted from \(t^{old}_{event}\) to \(t^{new}_{event}\) according to the new temperature. This procedure is done using the equation
\begin{equation}
t^{new}_{event} = (t^{old}_{event} - t) \frac{R_{hop}(T_{old}) + R_{eva}(T_{old})}{R_{hop}(T_{new}) + R_{eva}(T_{new})} + t.
\end{equation}
Since the ratio between $R_{hop}$ and $R_{eva}$ also changes, the type of event that occurs at $t^{new}_{event}$ must also be redetermined. 
 
\subsection{Cooling of the grain}
\label{cool}
The temperature of the grain is recalculated at certain time intervals using
\begin{equation}
\Delta t = - \int_{T_{new}}^{T_{old}} \frac{c(T)}{ \frac{\textrm{d}E}{\textrm{d}t}}\textrm{d}T,
\label{Dt}
\end{equation}
where \(\frac{\textrm{d}E}{\textrm{d}t}\) is given by
\begin{equation}
\frac{\textrm{d}E}{\textrm{d}t} = 4\pi \left(\pi a^2\right) \int_0^{\infty} Q_{em}\left(\lambda \right)B_\lambda \left(T \right) \textrm{d}\lambda
\label{dEdt}
\end{equation}
with \(Q_{em}\) the emission coefficient and \(B\) the Planck function.
We found that \(\Delta t(s) = \max\left(10r^2/R_{photon}, 30\right)\) with \(r\) in $\mu$m, is a good time interval, because it takes into account the difference in photon flux and the temperature fluctuations. 

We integrate Eq.~(\ref{dEdt}) numerically over the range 91.2 nm $< \lambda \leq$ 10 mm using \(Q_{em}\left(\lambda \right) = Q_{abs}\left(\lambda \right)\) and solve Eq.~(\ref{Dt}) to obtain the new temperature.

\begin{figure}
\begin{center}
\includegraphics[width=0.47\textwidth]{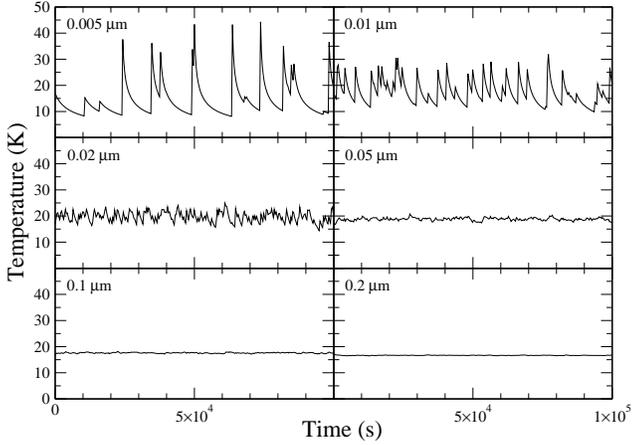}
\end{center}
\caption{The temperature fluctuations as a function of time for different grain sizes and $A_V$ = 0.}
\label{Temp}
\end{figure}

\section{Temperature fluctuations}
\label{tempfluc}
Before studying the formation of molecular hydrogen, we first discuss the influence of the incoming photons on the grain temperature. Figure \ref{Temp} shows how the grain temperature fluctuates as a function of time for different grain sizes at $A_V$ = 0. It clearly shows that although smaller grains have fewer incoming photons per given time interval, the temperature fluctuations are much larger, which is a consequence of the \(r^3\) term in the heat capacity. For a grain size (radius) of 0.005 \(\mu\)m,fluctuations of up to 30 K occur, while for a grain size of 0.05 \(\mu\)m, an almost constant temperature is obtained. Furthermore, it can be noticed that the minimum temperature reached before each photon hit is almost independent of the height of the temperature increase given the rapid initial cooling. \cite{Draine:2003} shows similar graphs for carbonaceous grains, in which the  temperature for the grains has similar fluctuations both in amplitude and in modal temperature.

\begin{figure}
\begin{center}
\includegraphics[width=0.47\textwidth]{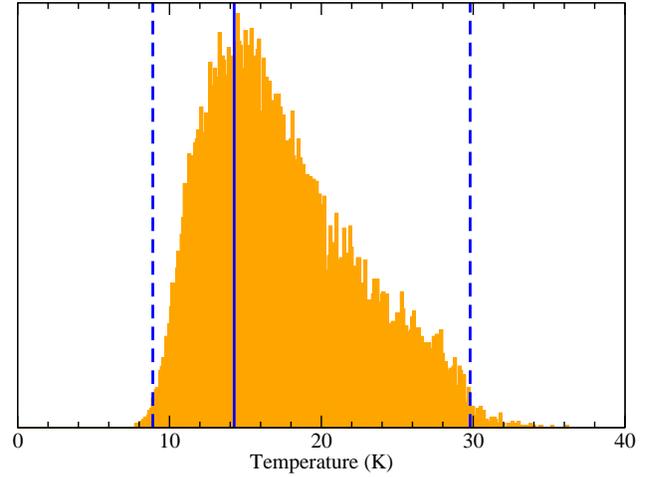}
\end{center}
\caption{The temperature distribution over time for \(A_V=0\) and \(r=0.01\) $\mu$m. The lines indicate the modal temperature (solid) and the two 99\% levels (dashed). See the text for more detail.}
\label{hist}
\end{figure}

\begin{figure}
\begin{center}
\includegraphics[width=0.47\textwidth]{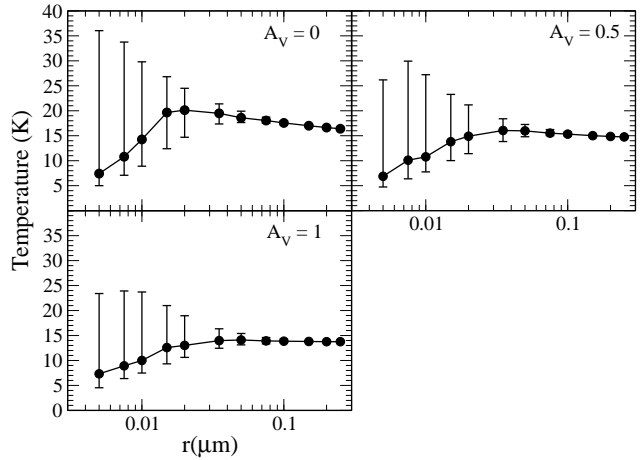}
\end{center}
\caption{Modal temperature of the grain as a function of grain size for different visual extinctions. The bars indicate 99\% confidence levels.}
\label{Tvsa}
\end{figure}

To obtain the modal grain temperature and the amplitude of the fluctuations, we determined the temperature at fixed intervals chosen such that there were several photon hits. Figure \ref{hist} shows a histogram of the temperature distribution so obtained for \(A_V=0\) and \(r=0.01\) $\mu$m. The histogram clearly has a maximum, at which the temperature of the grain occurs most frequently. This mode is indicated by a solid line, which is left of the centre of the distribution as could be expected considering the shape of the temperature vs. time peaks. Figure \ref{Tvsa} gives these modal grain temperatures as a function of grain size for \(A_V=0\), 0.5 and 1. The bars indicate the 99\% boundaries; i.~e.~the temperatures between which 99\% of the determinations fall. These boundaries are indicated in Figure \ref{hist} by dashed lines and are a measure of the amplitude of the fluctuations. Figure \ref{Tvsa} clearly shows that the temperature fluctuations are larger for smaller grain sizes. All three temperature profiles show a maximum, which is at \(r_m=0.02\) \(\mu\)m for \(A_V=0\) and gradually moves to larger \(r\) for higher \(A_V\). The maximum is sharper  at \(A_V \sim 0\), which indicates that the radiation causing this peak is in the UV  since this region is strongly shielded at higher extinction. The effect is due to the \(r\) dependence of the absorption coefficient in the high energy range, as discussed in Section \ref{rad field}.

\begin{figure}
\begin{center}
\includegraphics[width=0.47\textwidth]{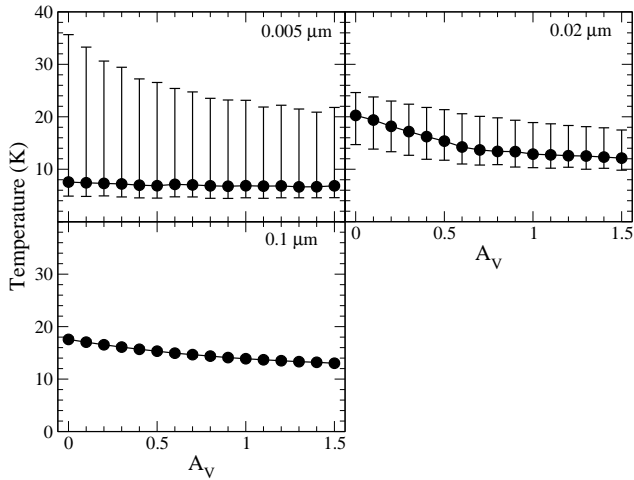}
\end{center}
\caption{Modal temperature of the grain as a function of \(A_V\) for different grain sizes.}
\label{TvsAv}
\end{figure}

\cite{Li:2001} show a similar graph, where the average temperatures of both graphite and silicate grains are plotted for different radiation field intensities. Comparison between the \(\chi_{MMP} = 1\) curve from that graph and the upper left panel of Figure \ref{Tvsa} shows that the shape of both curves is the same, but that the temperatures in Figure \ref{Tvsa} are slightly higher (1 K). This is a good agreement considering the difference in detail used in the methods. Note that in the limited regime depicted by \cite{Li:2001} ($r> 0.01$ $\mu$m) the modal and average temperatures coincide.

Figure \ref{TvsAv} gives the modal grain temperature but now as a function of \(A_V\) for three different grain sizes. The three curves show very different behaviour. The modal temperature of the smallest grain remains constant as a function of extinction but the amplitude of the fluctuations clearly decreases. The intermediate grain of radius 0.02 \(\mu\)m has a gradually decreasing modal temperature but the amplitude remains constant. It is interesting how the position of the mode in the distribution changes from the centre for \(A_v=0\) to the lower temperatures for \(A_v=1.5\). Finally, the 0.1 \(\mu\)m grain has almost no temperature fluctuations and the curve shows a gradual decrease of the temperature for grains deeper into the cloud. Notice the crossing of the curves for 0.02 \(\mu\)m and 0.1 \(\mu\)m around \(A_V=1\), which is a consequence of the gradual moving of \(r_m\) to larger values for higher visual extinction.

\section{Effect of the temperature on the efficiency}
\label{temp vs eff}
The previous section showed that not only the amplitude of the temperature fluctuations is grain-size dependent but also the modal temperature. This means that we should not limit our study to sizes where fluctuations play a role but study a wider range of grain sizes in order to obtain the complete size dependence of the recombination efficiency.

\begin{figure}
\begin{center}
\includegraphics[width=0.47\textwidth]{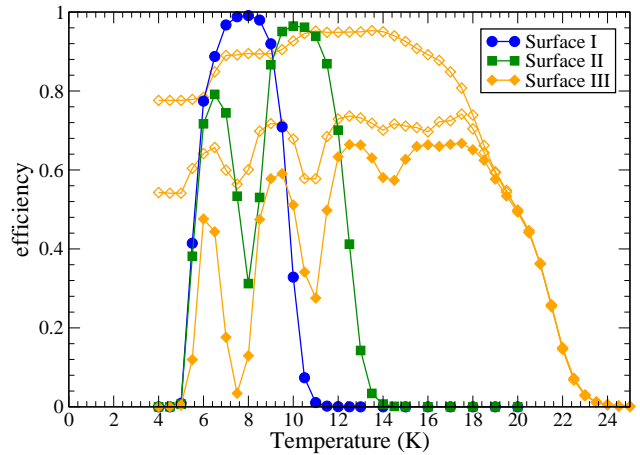}
\end{center}
\caption{The recombination efficiency as a function of temperature for the three different surfaces in the absence of fluctuations. The closed markers all indicate simulations with \(\mu=1\), while for Surface III, additional simulations are run and designated by open markers for \(\mu=0\) (upper curve) and \(\mu=0.3\) (lower curve).} 
\label{ethaCT}
\end{figure}

The definition of the recombination efficiency, $\eta$, used in this paper is
 \begin{equation}
 \eta = \frac{2 N_{\el{H}_2}}{N_{\el{H}}},
 \label{eta}
 \end{equation}
where \(N_{\el{H}}\) is the number of hydrogen atoms that approach the surface in a given time interval and \(N_{\el{H}_2}\) is the number of molecules that leave the surface in that interval. Usually, $\eta$ is defined as 
 \begin{equation}
 \eta = \frac{2 R_{\el{H}_2}}{F},
 \label{eta_rate}
 \end{equation}
where \(R_{\el{H}_2}\) is the rate with which H$_2$ comes off the surface and \(F\) is the flux of incoming hydrogen atoms. Since the Monte Carlo method uses discrete numbers and not rates, as in rate equation methods, the evaporation rate cannot be determined directly. We therefore use Eq.~(\ref{eta}) which gives the same results as Eq.~(\ref{eta_rate}) in steady-state conditions.

We first consider the efficiencies as a function of the temperature for the three different surfaces introduced in Section \ref{Model}. The results are shown in Figure \ref{ethaCT} and were obtained with a simulation in which temperature fluctuations were omitted. The size of the grains is large enough that the efficiency is roughly independent of it \citep{Chang:2005}, a regime known as the accretion limit. These graphs will help us to interpret the results we obtain from the simulations with temperature fluctuations. The efficiency curve for surface I, flat olivine, has one narrow peak at 8 K, surface II (olivine with islands) has one broad peak (7-11 K) punctuated by one dip and surface III (very rough surface) has a very broad peak (6-18 K) with three dips. These dips are due to hydrogen molecules, which occupy the high energy sites ($i>0$) and prevent hydrogen atoms from using these sites to make new molecules. At slightly higher temperatures the formed molecules evaporate and the sites are available again. Since we use \(\mu=1\) all molecules remain on the surface after formation until they evaporate. In our previous papers \citep{Chang:2005,Cuppen:H2} we used \(\mu=0\) and only one peak was found in all cases. The open markers in Figure \ref{ethaCT} give simulation results for \(\mu=0\) and \(\mu=0.3\) on Surface III, where the $\mu$ dependence is significant. The value of 0.3 is chosen since it is the value found by \cite{Katz:1999} for H\(_2\) formation on olivine. Notice that for certain temperatures the difference in efficiency can be as large as a factor of 10. These differences are however limited to very small temperature ranges so that the value of $\mu$ only influences the average efficiency over the range within a much smaller factor between the two extremes $\mu=0$ and $\mu=1$. The efficiency curves for $\mu=0.3$ and $\mu=1$ are much closer. We use \(\mu=1\) for now.

\begin{figure}
\begin{center}
\includegraphics[width=0.47\textwidth]{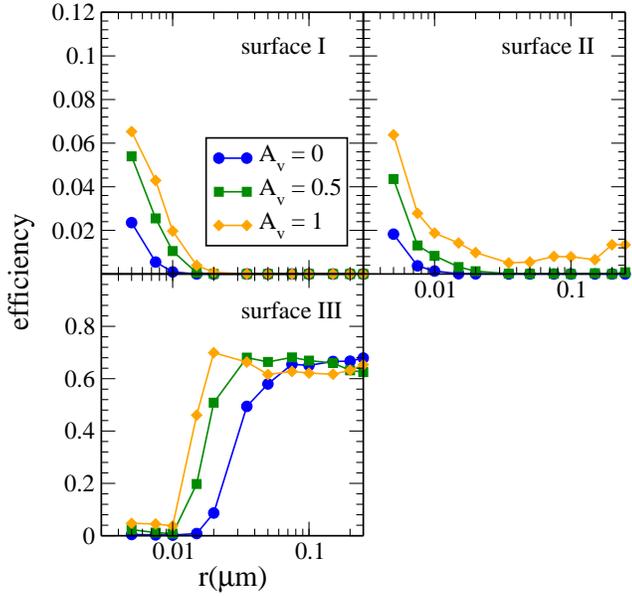}
\end{center}
\caption{The recombination efficiency as a function of the grain size for different surfaces and visual extinctions.}
\label{ethaSS3}
\end{figure}

Figure \ref{ethaSS3} shows the recombination efficiency as a function of grain size with the temperature fluctuations depicted as in Figure \ref{Tvsa} for the three different surfaces. With our assumed surface site density of \(2\times10^{14}\) cm\(^{-2}\), we use array sizes of \(25 \times 25\), 
\(50 \times 50\), and \(100 \times 100\) to represent the surfaces of grains of radii \(r\) = 0.005, 0.01, and 0.02 \(\mu\)m, respectively. The efficiency at a given temperature for larger grains is in all cases independent of size. So we can use a \(100 \times 100\) array for such grains as long as the temperature fluctuations are headed correctly.

Consider first the results for the smallest grains: \(r=0.005\) \(\mu\)m. From Figures \ref{Tvsa} and \ref{TvsAv} we know that these grains have a modal grain temperature of 8 K. According to Figure \ref{ethaCT} the efficiencies at this temperature are approximately 1.0, 0.3 and 0.1 for surfaces I, II, and III, respectively, without temperature fluctuations. Figure \ref{ethaSS3} shows however that for these small grains the recombination efficiency is similar for all surfaces and rather low. The temperature fluctuations, which occur frequently compared with the H-atom deposition interval, increase the evaporation rate which causes the average residence time of atoms on the surface to be reduced to such an extent that the efficiency is much lower and the atoms are no longer available when the grain cools down. As the grain size becomes somewhat larger, the modal temperature rises and the efficiency actually decreases for surfaces I and II, showing the rise in the modal temperature to outweigh the decrease in fluctuations.

If we look at the larger grains, \(r\geq0.05\) \(\mu\)m, the fluctuations become even less salient and the efficiencies are as could be expected combining Figures \ref{Tvsa} and \ref{ethaCT}. A grain of size \(r\geq0.1\) \(\mu\)m at \(A_V = 1\) for example has a temperature of 14 K which would result in efficiencies of approximately 0, 0, and 0.65 for surfaces I, II, and III, respectively, without temperature fluctuations. Figure \ref{ethaSS3} shows that this is indeed the case. 

\begin{figure}
\begin{center}
\includegraphics[width=0.47\textwidth]{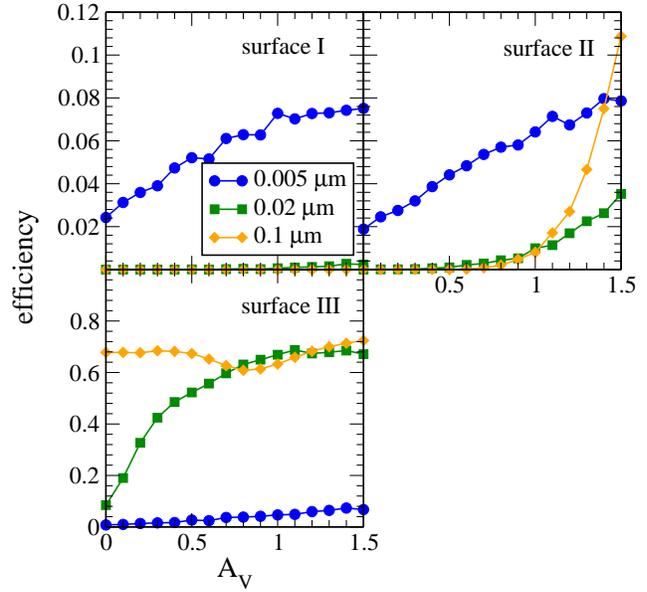}
\end{center}
\caption{The recombination efficiency as a function of \(A_V\) for different surfaces and grain sizes.}
\label{ethavsAv2}
\end{figure}

Figure \ref{ethavsAv2} shows the efficiency again, but now as a function of visual extinction for three grain sizes, and should be studied in combination with Figure \ref{TvsAv}. All graphs show an increase for increasing \(A_V\), except the \(0.1\) \(\mu\)m curve for surface III. These increases in efficiency are due to the temperature decrease as a function of visual extinction for grains of \(0.02\) \(\mu\)m and due to a decrease in fluctuations for grains of \(0.005\) \(\mu\)m. For grains of \(0.02\) \(\mu\)m  with surface III, the fluctuations gradually move inside the temperature range where a non-zero efficiency is obtained. This occurs at \(A_V\) = 0.7,  which is the reason that for higher extinction the efficiency hardly changes. The minimum in the \(0.1\) \( \mu\)m curve for surface III corresponds to the dip in the efficiency curve (Figure \ref{ethaCT}) around 15 K, since the temperature decreases from 17.5 K to 13 K going from \(A_V=0\) to 1.5. 

\begin{figure}
\begin{center}
\includegraphics[width=0.47\textwidth]{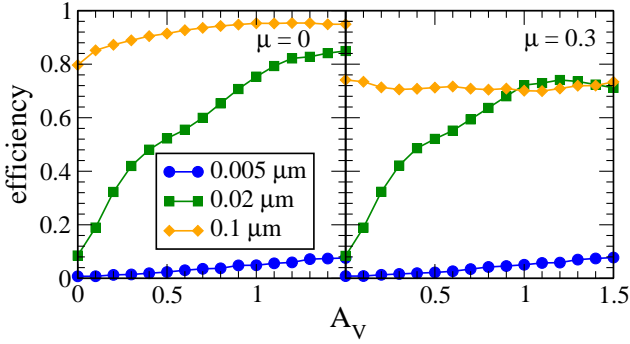}
\end{center}
\caption{The efficiency as a function of the visual extinction for different values of \(\mu\) for surface III and different grain sizes.}
\label{ethavsmu}
\end{figure}

Figure \ref{ethavsmu} shows the efficiency as a function of \(A_V\) for \(\mu=0\) and 0.3 for surface III; it can be compared with the lower left panel of Figure \ref{ethavsAv2}, which gives results for \(\mu=1\).  We have showed already that the value of $\mu$ can have a strong effect on the efficiency at specific small temperature ranges, but have also mentioned that on average the effect is much smaller. The low efficiency ranges with $\mu=1$ are around 8 and 11 K, at which newly formed H$_{2}$ molecules interfere with subsequent reactions. Although the grains of \(0.005\) \(\mu\)m have a modal temperature of 8 K, the fluctuations cause H\(_2\) to evaporate easily, resulting in a low effect of \(\mu\) on the efficiency. For the other two grain sizes we see a visible but still small effect due to variations in \(\mu\). This could be expected considering that the modal temperatures (12-20 K) do not fall in the low efficiency dips. 

\section{Consequences for the diffuse ISM}
\label{ISM}
In the previous sections, we showed that the amplitude of the temperature fluctuations and the modal temperature of grains are strongly size dependent, and that both quantities have a strong effect on the efficiency of molecular hydrogen formation on olivine. While for the large grains the type of surface also has a strong effect on the efficiency, the roughness of the surface seems to have little effect for the smallest grains, since for those grains there is enough thermal energy for atoms to hop and evaporate even from the high energy sites because of the strong temperature fluctuations. 

This section discusses the consequences of these findings for molecular hydrogen formation in the diffuse interstellar medium. Usually, formation rates for molecular hydrogen are based on the assumption that all grains have a standard size of 0.1 \(\mu\)m. In reality, there is a wide range of different sizes in the ISM. The distribution of these sizes highly favours small grains and can be approximated by a power law \citep{Mathis:1977}:
\begin{equation}
n_g(r) = cr^{-3.5},
\label{n_g}
\end{equation}
{where $n_g(r)\textrm{d}r$ is the concentration of grains with radii in the range ($r$, $r+\textrm{d}r$ in cm$^{-3}$).}
{\cite{Draine:1984} found a value  \(c=1.51\times 10^{-25}n_H\) cm$^{-\frac{1}{2}}$ for the proportionality constant using \(r_{min}=0.005\) \(\mu\)m and \(r_{max}=0.25\) \(\mu\)m \citep{Mathis:1977} for the minimum and maximum radii. The gas-to-dust number ratio is found by integrating Eq.~(\ref{n_g}) over the range and dividing by \(n_H\) to be \(2.91 \times 10^{9}\), which is significantly lower than the usually-assumed value of 10\(^{12}\), a number that only considers 0.1 \(\mu\)m grains and therefore neglects the large portion of small grains. A somewhat higher value of \(6.0 \times 10^{9}\) was obtained by \cite{Lipshtat:2004} who only considered grains down to 0.04 $\mu$m in size. 
}

From the grain size distribution we can calculate the rate coefficient $R$ in cm$^{3}$ s$^{-1}$ for molecular hydrogen formation:
\begin{equation}
R = \frac{\int_{r_{min}}^{r_{max}} \frac{1}{2} n_g(r) v \eta(r) \xi \pi r^2 \textrm{d}r}{n_{\el{H}}},
\label{R}
\end{equation}
where \(\xi\) is the sticking coefficient of the atoms to the grain. For the low temperatures used here the sticking coefficient can be assumed to be close to one \citep{Buch:1991,Al-Halabi:2002}. Eq.~(\ref{R}) can be simplified to 
\begin{equation}
R = \frac{\alpha}{n_H} \sqrt{\frac{T_{gas}}{300}},
\end{equation}
where 
\begin{equation}
\alpha = c \sqrt{\frac{600\pi k}{ m}}\int_{r_{min}}^{r_{max}}  \eta(r) r^{-1.5} \textrm{d}r.
\label{alpha}
\end{equation}
Note that in the standard treatment where all grains are 0.1 $\mu$m in size and $\eta=1$, $\alpha = 5\times 10^{-17}$ cm$^{3}$ s$^{-1}$. This is in agreement with the estimate of $R = 10^{-17}$ cm$^{3}$ s$^{-1}$ for diffuse-cloud conditions given by \cite{Jura:1974}, which corresponds with  $\alpha \approx 2\times10^{-17}$ cm$^{3}$ s$^{-1}$.

\begin{figure}
\begin{center}
\includegraphics[width=0.47\textwidth]{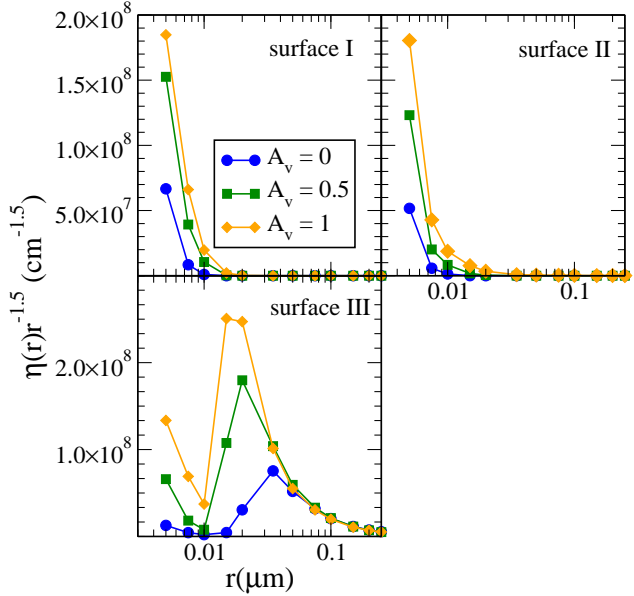}
\end{center}
\caption{The recombination efficiency $\times$ \(r^{-1.5}\) (or integrand in Eq.~(\ref{alpha})) as a function of the grain size for different surfaces and visual extinctions.}
\label{ethaxr}
\end{figure}

Figure \ref{ethaxr} shows \(\eta(r) r^{-1.5}\) as a function of the grain size using the data from Figure \ref{ethaSS3}. For surfaces I and II the maxima for the efficiency and \(r^{-1.5}\) coincide at $r_{min}$, which obviously leads to a maximum for the smallest grain sizes used. Surface III, however, has its largest efficiencies for the largest grains. This results in a minimum for \(\eta(r) r^{-1.5}\) around $r=0.01$ $\mu$m and a maximum at intermediate grain sizes, depending on $A_V$. Figure \ref{ethaxr} clearly shows that the higher efficiency for large grains is of almost no importance. For flat grains and those with a minimal roughness, the overall H$_2$ formation is mainly determined on grains with  $r\leq 0.01$ $\mu$m, while for grains with a high roughness, the intermediate range of $0.01\leq r\leq 0.08$ $\mu$m is most important. Since surfaces I and II have their maximum value for \(\eta(r) r^{-1.5}\) at \(r_{min}\), the choice for the lower limit in grain size distribution influences the value for \(\alpha\). Generally the minimum size for olivine grains is considered to be $r=0.005$ $\mu$m \citep{Mathis:1977,Draine:1984}, which is the value we take. {\cite{Li:2001a} concluded, however, that up to $\sim10\%$ of interstellar Si could be in $r\leq 15$ \AA ~silicate grains. We did not consider these ultra-small particles in our calculations because, in our view,  such particles need to be treated as a special case for several reasons: (i) they consist of up to only a few hundred molecular units and therefore may not retain their bulk properties, such as the heat capacity that we use to calculate the heating and cooling; (ii) the small number of units on the surface means that our smooth and rough topologies are imprecise, (iii) the efficiency of sticking by weakly bound adsorbates is unlikely to be high \citep{Herbst:1991} and it is likely that chemisorption is needed for high sticking efficiencies; and (iv) strong adsorbate-substrate bonds may be necessary for H atoms to withstand the very large temperature pulses that photons will cause.  Such considerations will also be necessary to treat molecular hydrogen formation on PAH's. }  

Table \ref{Tab_R} shows our numerically integrated $\alpha$ coefficients. Note that the efficiencies used for the integration are only determined for one gas temperature and one gas concentration of H. We did not check for an additional dependence of the efficiency on these two quantities. Considering that the efficiencies used to obtain the values in Table \ref{Tab_R} are much lower than unity, the \(\alpha\) coefficients are surprisingly close to \(5 \times 10^{-17}\) cm$^{3}$ s$^{-1}$. But we must remember that the large number of small grains gives a very large surface area for reaction. Indeed, if we consider a constant high efficiency of one over the whole grain size range, an $\alpha$ coefficient of \(1.53\times10^{-16}\) cm$^{3}$ s$^{-1}$ is found. Of course our computed values are lower.  From Table \ref{Tab_R}, we can see that only our rough surface III is able to produce a sufficient amount of molecular hydrogen at $A_V=0$, while the other surfaces are possible alternatives, with $\alpha$ a  factor of {six} below Jura's value, at $A_V=1$, which pertains at more internal regions for diffuse clouds. This difference is caused by the intermediate grains, which have a very low efficiency for surface I and II, but have appreciable values for surface III.

The low efficiency of surfaces I and II for intermediate grains is due ultimately to the low evaporation energy for H given by \cite{Katz:1999}. In their paper they do indicate that their stated energy is more precisely only a lower limit {within a range of 3 meV}. Using a higher value results in larger $\alpha$ coefficients for surfaces I and II since H evaporates more slowly, and makes these surfaces more suitable candidates for efficient molecular hydrogen formation, especially at non-zero extinction.
{The last row in Table \ref{Tab_R} gives the $\alpha$ coefficients for surface II if the value of $E_D$ is increased by the extra 3 meV (35 K, an increase of 9\%.) We see that there is a clear increase in the $\alpha$ coefficient of a factor of a few for low $A_V$ that gets larger as the extinction increases, so that for $A_V>0.5$ the rate coefficient is high enough to produce a sufficient amount of molecular hydrogen in diffuse clouds.}

\begin{table}
\caption{$\alpha$ coefficient in cm$^{3}$ s$^{-1}$.}
\begin{tabular}{lccc}
\hline
               & \(A_V=0\)            & \(A_V=0.5\)          & \(A_V=1\)\\
\hline
surface I      & $6.50\times10^{-19}$ & $1.99\times10^{-18}$ & $2.91\times10^{-18}$\\
surface II     & $5.08\times10^{-19}$ & $1.51\times10^{-18}$ & $3.20\times10^{-18}$\\
surface III    & $3.12\times10^{-17}$ & $4.75\times10^{-17}$ & $5.62\times10^{-17}$\\
surface II$^1$ & $1.32\times10^{-18}$ & $5.42\times10^{-18}$ & $2.74\times10^{-17}$\\
\hline

\end{tabular}
\\
$^1$ $E_D$ is increased by 35 K.
\label{Tab_R}
\end{table}

Table \ref{Tab_R_c} gives values of the $\alpha$ coefficient for simulations performed at a constant temperature. We choose the modal temperatures of the standard grain of \(r=0.1\) $\mu$m, which are 17.6, 15.3, and 13.9 K for \(A_V=0\), 0.5 and 1, respectively. In this way, we can separate out the effects of the low modal temperature and stochastic heating on small grains. No H$_2$ was formed within our simulation time for the completely flat surface I and slightly rough surface II at $A_V=0$; we can therefore only give an upper limit for these cases. The higher values in the case of the stochastic heating (Table \ref{Tab_R}) stem from the lower modal temperature for the small grains. Surface III, on the other hand, has a small increase in the formation rate at constant temperature due to the higher efficiency for small grains in the absence of the fluctuation in temperature. Here the increase in modal temperature is not as important because of the large temperature range for high efficiency.

\begin{table}
\caption{$\alpha$ coefficient in cm$^{3}$ s$^{-1}$ at a constant temperature.}
\begin{tabular}{lccc}
\hline
               & \(A_V=0\)               & \(A_V=0.5\)             & \(A_V=1\)\\
\hline
surface I      & $<$$1.72\times10^{-22}$ & $<$$1.72\times10^{-22}$ & $<$$3.82\times10^{-22}$\\
surface II     & $<$$2.85\times10^{-22}$ & $3.28\times10^{-20}$    & $1.41\times10^{-18}$\\
surface III    & $1.03\times10^{-16}$    & $1.04\times10^{-16}$    & $9.75\times10^{-17}$\\
\hline
\end{tabular}
\label{Tab_R_c}
\end{table}

\section{Conclusions}
\label{conclusions}
Previous calculations of the rate of molecular hydrogen formation on interstellar grain surfaces in diffuse clouds have ignored the subtle role played by radiation in determining the surface temperature and its fluctuations as functions of grain size.  As shown by earlier investigators (e.g. \cite{Draine:2001}) and confirmed here, the temperature of grains fluctuates strongly for smaller grains ($r < 0.02$ $\mu$m) but not for larger ones, and the modal temperature peaks for grains in the middle range of size (0.02 - 0.05 $\mu$m).  
We have shown that the efficiency of formation of H$_2$ depends both on the modal surface temperature and its fluctuations.  According to the type of material and its smoothness or roughness, each surface has a range of temperature where the efficiency is high.  For olivine grains, which are considered here, a smooth surface has only a very small range of surface temperature for high efficiency  6-9 K - while a very rough surface, with half of its binding sites different from the norm, allows efficient H$_2$ production up to about 20 K.  For small grains, photons pulse the temperature above the range of efficiency, thus reducing it, although the low modal temperature aids H$_2$ production, especially for flat and moderately rough olivine grains. The net result for these two types of surfaces is an enhancement in the efficiency for smaller vs. larger grains although the overall efficiency remains low at all sizes.  The relatively high efficiency for small grains is augmented by the high overall surface area given the distribution of grain sizes, so that the production of H$_2$ is dominated by the smaller grains.  For very rough grains, on the other hand, the efficiency is greater for larger grains since fluctuations are minimal and the modal temperature is still within the range of high efficiency, while it is above this range for smoother surfaces.  Here the countervailing larger surface area for smaller grains results in a peak H$_2$ production for grains of intermediate sizes. 

A calculation of the rate coefficient $R$ for hydrogen formation in diffuse clouds for which grains from a minimum radius of 0.005 $\mu$m to a maximum radius of 0.25 $\mu$m are included shows that in order to equal or exceed the value of $R$ deduced by \cite{Jura:1974} to reproduce the H$_2$/H balance in diffuse clouds, one needs a very rough surface of olivine if $A_V=0$.  At a slightly higher visual extinction of 1.0, flat and moderately rough surfaces lead to values of $R$ a factor of {six} lower than the \cite{Jura:1974} value.  Since the latter visual extinction is quite reasonable for internal portions of  diffuse interstellar clouds, it appears that the rate of formation of H$_2$ on olivine grains of widely differing topologies can possibly account for the production of this species although the rougher surfaces are better candidates.  This conclusion rests ultimately on the energy needed for evaporation of H atoms deduced by \cite{Katz:1999} for an olivine surface; a somewhat higher value, also consistent with their data, improves the suitability of the smoother surfaces for efficient hydrogen production, especially if the extinction is greater than zero. If the effects of the radiation field are removed, on the other hand, the difference between the smoother and rougher surfaces remains large.

In our calculations, we have only considered grains {larger than 5 nm in radius} consisting of olivine,  and we have only treated a standard radiation field.  {The inclusion of smaller silicate particles is an obvious extension, as is} the inclusion of carbonaceous grains, which would also allow us to consider much smaller particles, including PAHs.  For these very small particles, temperature fluctuations will be severe, and strong H atom-surface bonds, known as chemisorption \citep{Cazaux:2004}, must probably be considered to allow H atoms to remain on the grains at the peak temperatures. It is possible, if barriers to chemisorption are low, that the inclusion of { small silicate particles} and PAHs will increase the calculated values of the rate coefficient $R$ for diffuse clouds, and we intend to investigate this effect in the near future.  
On the other hand, an enhanced radiation field, as associated with photon-dominated regions, will increase the frequency of fluctuations in temperature for small grains, and likely decrease the overall rate of H$_2$ formation although the effect has yet to be studied. 

\section{Acknowledgements}

E. H. wishes to thank the National Science Foundation (U. S.) for support of his research programme in astrochemistry. {The authors wish to thank the referee for his careful reading of the manuscript.}

\end{document}